\documentclass[twocolumn,footinbib,showpacs,reprint,aip, apl]{revtex4-1} 
\usepackage{color}
\usepackage{amsfonts,amssymb,amsmath,mathrsfs}
\usepackage{amsbsy}
\usepackage{graphicx}
\usepackage{hyperref}

\makeatletter

\begin{document}

\title{Magnetic skyrmion bubble motion driven by surface acoustic waves}

\author{Rabindra Nepal}
\affiliation{Department of Physics and Astronomy and Nebraska Center for Materials and Nanoscience, University of Nebraska, Lincoln, Nebraska 68588, USA}

\author{Utkan G\"ung\"ord\"u}
\affiliation{Department of Physics and Astronomy and Nebraska Center for Materials and Nanoscience, University of Nebraska, Lincoln, Nebraska 68588, USA}

\author{Alexey A. Kovalev}

\affiliation{Department of Physics and Astronomy and Nebraska Center for Materials and Nanoscience, University of Nebraska, Lincoln, Nebraska 68588, USA}

\begin{abstract}
We study the dynamical control of a magnetic skyrmion bubble by using counter-propagating surface acoustic waves (SAWs) in a ferromagnet. First, we determine the bubble mass and derive the force due to SAWs acting on a magnetic bubble using Thiele's method. The force that pushes the bubble is proportional to the strain gradient for the major strain component. We then study the dynamical pinning and motion of magnetic bubbles by SAWs in a nanowire. In a disk geometry, we propose a SAWs-driven skyrmion bubble oscillator with two resonant frequencies. 
\end{abstract}

\maketitle

Efficient manipulation of magnetic textures such as domain walls and skyrmions is a long-standing quest in spintronics research. \cite{Boulle.Malinowski.ea:MSaER2011, Iwasaki.Mochizuki.ea:NC2013} It is attracting a lot of interest due to its potential applications in magnetic memory devices. \cite{Parkin.Hayashi.ea:S2008, Yang.Ryu.ea:NN2015, Fert.Cros.ea:NN2013} One of the major obstacles in such studies is the large current or magnetic field required to drive the system. \cite{Beach.Tsoi.ea:JMMM2008} The presence of pinning sites will further hinder the efficient manipulation, \cite{Bogart.Atkinson.ea:PRB2009, Jiang.Thomas.ea:NL2011, Reichhardt.Ray.ea:PRL2015} however, skyrmions and topological bubbles in general require much lower depinning currents. In addition to longitudinal motion, skyrmions also show transverse Hall-like motion, \cite{Nagaosa2013, PhysRevLett.111.067203, PhysRevLett.112.187203, PhysRevLett.107.136804,Kovalev:PRB2014,Woo2015, Jiang2016a,Litzius.Lemesh.ea:NP2017} which in some cases can complicate device realizations.\cite{Liang.Degrave.ea:NC2015} Alternative methods relying on conservation of mechanical angular momentum can also be used for magnetization control.\cite{PhysRevApplied.5.031002, PhysRevLett.117.237201}

To avoid dissipation associated with transport one can employ electric field control of magnetization. Such control can be realized via various magnetoelectric effects \cite{Tong.Fang.ea:CMS2016, Hsu2016} and it has been demonstrated theoretically. \cite{Upadhyaya.Dusad.ea:PRB2013, Belashchenko.Tchernyshyov.ea:APL2016}
Since the modulation of anisotropy by strain is observed in many ferromagnetic materials, strain can also be used to control magnetization dynamics. \cite{Davis2010,Lei.Devolder.ea:NC2013,Dean.Bryan.ea:APL2015, Edrington2018, Li2017, Liang.Sepulveda.ea:JAP2015,Rousseau.Weil.ea:SR2016} Thus, one can also electrically control magnetization dynamics by combining piezoelectric and magnetoelastic effects. A possibility to drive a domain wall by electrically-induced surface acoustic waves (SAWs) has been demonstrated recently. \cite{Dean.Bryan.ea:APL2015, Edrington2018} Such studies pave the way for various applications in magnetic memory and logic devices. \cite{Lei.Devolder.ea:NC2013,DSouza.SalehiFashami.ea:NL2016, Zhang2015}

In this work, we explore a skyrmion bubble dynamics induced by counter-propagating SAWs. To properly describe the dynamics we introduce a finite skyrmion bubble mass. It has been established recently that the presence of finite mass can lead to unusual dynamics for the field and current induced skyrmion bubble motion. \cite{Makhfudz.Krueger.ea:PRL2012, Moon.Chun.ea:PRB2014, Buettner.Moutafis.ea:NP2015} We find that, in some instances, this mass substantially modifies the SAW induced motion of a skyrmion bubble.

We consider a magnetic skyrmion bubble, in a ferromagnetic nanowire with perpendicular magnetic anisotropy. The free energy density of the system well below the Curie temperature can be written as,
\begin{align}
\mathcal F_0 =& \frac{A}{2}(\partial_i \boldsymbol n)^2 - K_u n_z^2   + \frac{1}{2} \mu_0 M_s \boldsymbol H_d \cdot \boldsymbol n  + \mu_0 M_s  H n_z
\end{align}
where $i = \{x, y\}$, $\boldsymbol n$ is a unit vector in the direction of local spin density, $A$ is the exchange stiffness constant, $K_u$ is uniaxial anisotropy along the $z$-direction, $H$ is the magnetic field applied along the $z$-axis and is required to stabilize and control the size of the bubble, $M_s$ is the saturation magnetization and $\boldsymbol H_d=\boldsymbol H_d[M_s\boldsymbol n]$ is demagnetizing field. 

\begin{figure}[htb!]
\vspace{0.5cm}
\centering
\includegraphics[width=0.9\columnwidth]{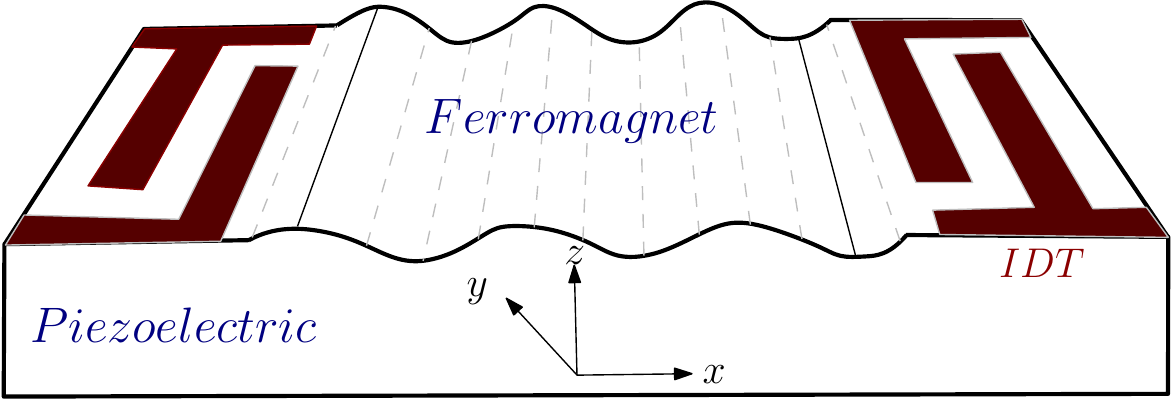}
\caption{Schematics of system setup: Counter-propagating SAWs applied through IDTs create surface 
displacements on the ferromagnetic layer that translate into strain components. }
\label{setup}
\end{figure}

We are interested in the dynamics of a magnetic bubble due to counter-propagating SAWs, see Fig.(\ref{setup}). Such SAWs can be generated using interdigitated transducers (IDTs) and form a standing wave along the sample provided that the two super-imposing waves are of same frequency. The magnetoelastic energy density for a film with cubic symmetry can be written as,
\begin{align}
\mathcal{F_{\text{ME}}} = B_1 \sum_{i}\epsilon_{ii}(z,t)n_i^2 + B_2 \sum_{j\ne i}\epsilon_{ij}(z,t)n_i n_j,
\label{fme}
\end{align}
where $B_1$ and $B_2$ are the magnetoelastic constants of the ferromagnetic material, and $\epsilon_{ij}$ stands for various components of the strain tensor. A SAW in a general case of an anisotropic piezoelectric can be associated with the strain components $\epsilon_{xx}$, $\epsilon_{xy}$, $\epsilon_{xz}$, and $\epsilon_{zz}$. \cite{Campbell.Jones:IToSaU1968} We use the following parametrization for the strain components, 
\begin{equation}
\begin{aligned}
\epsilon_{ij} = \varepsilon_{ij}^0\Big( \sin(k x + \omega_{+} t + \Phi_{ij}^R) + \sin(k x - \omega_{-} t + \Phi_{ij}^L) \Big)
\end{aligned}
\end{equation}
where $\Phi_{ij}^R$ and $\Phi_{ij}^L$ describe the phase differences of the SAWs propagating in the opposite directions, $\varepsilon_{ij}^0$ is the strain amplitude, $k$ and $\omega_{\pm}$ are the wavevector and frequencies of the strain waves, here $\omega_{\pm} = \omega \pm \Delta \omega$ with  $\Delta \omega$ being detuning between the counter-propagating waves. The total energy density we consider is $\mathcal{F} = \mathcal{F}_0 + \mathcal{F}_\text{ME}$.

We study the low energy dynamics of topological defects by employing the collective coordinates approach. \cite{Thiele1973} Within this approach the skyrmion bubble is treated as a particle experiencing a force due to magnetoelectric coupling in Eq.~(\ref{fme}).  A rotationally symmetric magnetization texture, $\boldsymbol n = (\sin \theta \cos \phi, \sin \theta \sin \phi, \cos \phi)$,  along the $z$-axis with helicity $\chi$ can be represented in cylindrical  coordinates as $n_\phi = W \varphi + \chi$ and $n_\theta = \theta(r, \varphi, z)$ where $W = \frac{1}{2\pi}\int_{\varphi=0}^{2\pi} d\phi$ is the winding number and $\theta(r, \varphi, z)$ is determined by minimizing the free energy. 
We assume that a magnetic bubble is described by the ansatz, \cite{Malozemoff} 
\begin{align}
\label{ansatz}
\theta(r, \varphi, z) = \pm 2 \arctan\Big[\exp \Big(\frac{P(r-R)}{\Delta}\Big) \Big]
\end{align}
where $P = \frac{1}{\pi} \int_0^\infty \frac{\partial \theta}{\partial r}dr = \pm 1$ is the polarity of the magnetic bubble, $R$ is the radius, and $\Delta$ is the width of the circular domain wall. 
We substitute this ansatz into the Landau-Lifshitz-Gilbert (LLG) equation,
\begin{align}
s(1+\alpha\boldsymbol n \times) \dot{\boldsymbol n} + \boldsymbol n \times \delta_{\boldsymbol n} F = 0,
\end{align}
where $F$ is the total free energy, $s=M_s / \gamma$ is the spin angular momentum density and $\gamma$ is the gyromagnetic ratio. By multiplying the LLG equation with $\frac{1}{4\pi}\int d^2 r \partial_{X,Y} \boldsymbol n \cdot \boldsymbol n \times$ and carrying out the integration, we obtain the equation of motion for the center of mass of the bubble $\boldsymbol R = (X, Y)$ as,
\begin{align}
4\pi s \left(\alpha \hat \eta-Q \boldsymbol e_z \times    \right) \dot{ \boldsymbol R}  + \boldsymbol F(\boldsymbol R) = 0.
\label{eq:Thiele}
\end{align}
Here $Q = (1/4\pi)\int dx dy ( \boldsymbol n \cdot \partial_x \boldsymbol n \times \partial_y \boldsymbol n) \equiv W P$ is the topological charge of the bubble, also known as skyrmion number, $[\hat\eta]_{ij} =\int d^2r \partial_i \boldsymbol n \cdot \partial_j \boldsymbol n/4\pi$ is the dyadic dissipation tensor and $\boldsymbol F = -\boldsymbol e_i \int \delta_{\boldsymbol n} \mathcal{F} \cdot \partial_{X,Y} \boldsymbol n = -\boldsymbol e_i \partial_{X,Y} F$ is the effective force (per unit thickness) due to strain on the bubble where $\boldsymbol e_i$ is a unit vector along the $i$-axis.
For a symmetric bubble, $\hat \eta$ is reduced to $\delta_{ij} \int d r [(r\partial_r n_\theta)^2 + \sin^2 n_\theta] /4r$. The rotational term in Eq.(\ref{eq:Thiele}) represents the Magnus force and pushes magnetization texture in transverse direction. With the assumption that magnetization texture under consideration moves as a rigid body without any deformation, the effective force along the $x$-direction on the texture due to SAWs is obtained as, 
\begin{align}
F_x(x = X) &= \frac{1}{k^2}\Big(A_{xx} \nabla \epsilon_{xx} + A_{zz} \nabla \epsilon_{zz} + A_{xy} \nabla \epsilon_{xy} \nonumber  \\
& + k A_{xz} \epsilon_{xz}\Big),
\label{force}
\end{align}
whereas the force along the $y$-direction is zero i.e. $F_y(X) = 0$, for the SAWs applied longitudinally.
The constants $A_{ij}$ in Eq.(\ref{force}) are the shape factors due to strain and their values depend upon the shape and size of the bubble. For a rotationally symmetric magnetization texture, $\theta(r, \varphi, z) \equiv \theta(r)$, the shape factors are given by,
\begin{widetext}
\begin{subequations}
\label{shapefactors}
\begin{eqnarray}
A_{xx} &=&- \pi B_1 \cos2\chi \int_{0}^{\infty} d\tilde{r} \frac{1}{\tilde{r}}\Big[4 J_2(\tilde{r}) \sin^2\theta + \tilde{r}\Big(2 J_2(\tilde{r}) - \tilde{r} J_1(\tilde{r}) \Big)\sin2\theta \partial_{\tilde{r}} \theta   \Big] + \pi B_1 \int_{0}^\infty d\tilde{r} \tilde{r} J_1(\tilde{r}) \sin2\theta \partial_{\tilde{r}}\theta, \label{equationa}
\\
A_{xy} &=& - \pi B_2  \sin 2 \chi \int_{0}^{\infty} d\tilde{r} \frac{1}{\tilde{r}} \Big[ 4 J_2(\tilde{r})  \sin^2\theta + \tilde{r}\Big( \tilde{r} J_1(\tilde{r}) - 2 J_2(\tilde{r})  \Big)\sin2\theta \partial_{\tilde{r}}\theta \Big], \label{equationb}
\\
A_{xz} &=& - \pi B_2 \cos \chi \int_{0}^{\infty} d\tilde{r} \frac{1}{\tilde{r}} \Big[  J_1(\tilde{r})  \sin2\theta + 2\tilde{r}\Big(J_1(\tilde{r}) - \tilde{r} J_2(\tilde{r})  \Big)\cos2\theta \partial_{\tilde{r}}\theta \Big], \label{equationc}
\\
A_{zz} &=& 2\pi B_1 \int_{0}^\infty d\tilde{r} \tilde{r} J_1(\tilde{r}) \sin2\theta \partial_{\tilde{r}}\theta,
\end{eqnarray}
\end{subequations}
\end{widetext}
where $\tilde{r} = k r$ and $J_n(\tilde{r})$ is the Bessel function of order $n$. \cite{Abramowitz} From Eq.~(\ref{equationb}), it is clear that a magnetic bubble or skyrmion with helicity $\chi = 0, \pm \pi/2$ or $\pm \pi $ doesn't couple with $xy$-component of strain but we observed coupling with $xy$-component when $\chi = \pi/4$. In our discussion we take $\chi = 0$ and concentrate on $\epsilon_{xx}$ component of the strain since for a thin magnetic layer all other strain components are small and have very little effect on the skyrmion bubble dynamics as has been confirmed by the numerical calculations.

\begin{figure*}
\centering
\includegraphics[width=2\columnwidth]{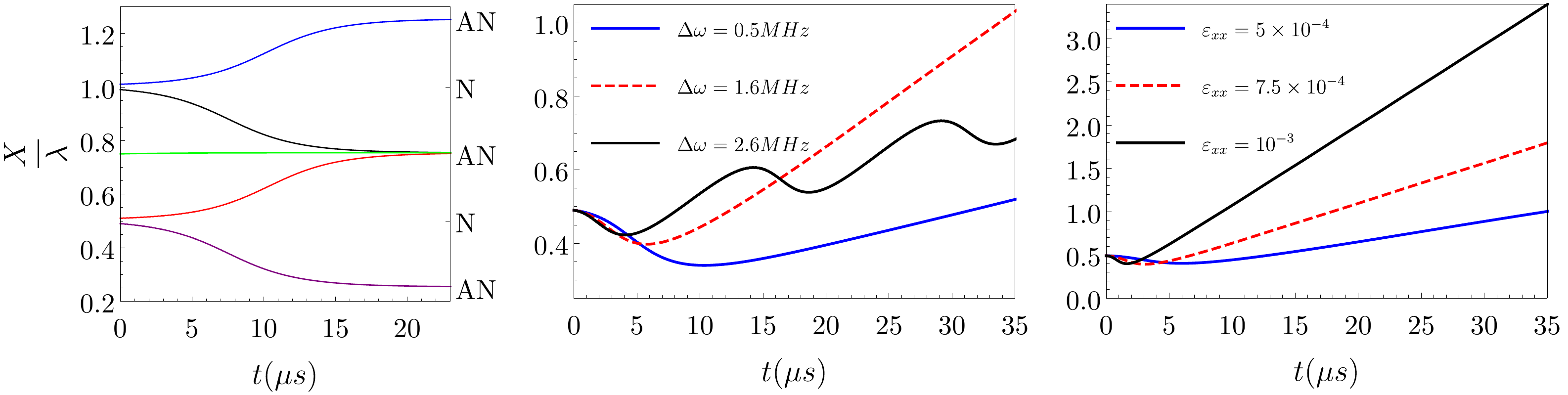}
\caption{(Color online) (Left) The translations of a magnetic bubble from nodes to antinodes of a standing wave. Magnetic bubble initially placed at different positions of the standing wave travels towards the nearest antinodes. (Middle) The longitudinal displacements of a bubble at different SAWs detunings.  (Right) Displacements of the bubble at different strain strengths.  Strain magnitude $\varepsilon_{xx}^0 = 5 \times 10^{-4}$ is used unless specified. }
\label{dynamics}
\end{figure*}

An evolving magnetic bubble gains inertial mass from the D\"oring mass due to the local magnetostatic energies and breathing modes. Since mass is an important factor in a rigid body dynamics, we phenomenologically introduce the bubble mass into Eq.~(\ref{eq:Thiele}) as,
\begin{align}
-\mathcal{M} \ddot{\boldsymbol R} + 4\pi s(\alpha \hat{\eta} - Q \boldsymbol e_z \times ) \dot{\boldsymbol R} - \boldsymbol \nabla U(\boldsymbol R)+  \boldsymbol F(\boldsymbol R,t) = 0.
\label{eoms}
\end{align}
where $\mathcal{M} = M/t$ with the mass of the bubble $M$  and thickness of the ferromagnetic sample $t$. And $\boldsymbol F(\boldsymbol R,t)$ is the strain force on the bubble induced by the strain. In addition, a bubble can experience repulsive force from the edges. Such force is introduced in Eq.~(\ref{eoms}) through the corresponding edge potential, i.e., a force given by $- \boldsymbol \nabla U(\boldsymbol R)$ for the edge potential $U(\boldsymbol R)$.
For example in a disk, a magnetic bubble moves inside a parabolic potential, $U(X, Y) = \mathcal{K} (X^2 + Y^2)/2$, where $\mathcal{K}$ is the spring constant corresponding to the repulsion of a bubble from the edges. 
Note that without the gyrotropic term the equation of motion of a magnetic bubble obtained above is similar to that of a damped harmonic oscillator driven by an external force.

We now study the dynamics of a magnetic bubble by numerically solving Eq.~(\ref{eoms}). In  the absence of strain force and damping, in a disk geometry with parabolic edge potential $U(X,Y)$, the magnetic bubble moves in a trajectory with  two circular modes of frequencies given by, \cite{Makhfudz.Krueger.ea:PRL2012}
\begin{align}
\label{frequencies}
\omega_{\pm} = -\frac{2 \pi s Q}{\mathcal{M}} \pm \sqrt{(2 \pi sQ/\mathcal{M})^2 + \mathcal{K}/\mathcal{M}}.
\end{align}
With the help of the open source micromagnetic simulator mumax3, \cite{Vansteenkiste2014} we first estimate the frequencies of the two circular modes of bubble motion in a disk of radius $100\text{nm}$ using the method similar to Ref.~\onlinecite{Makhfudz.Krueger.ea:PRL2012}. For a skyrmion bubble of radius $52\text{nm}$ stabilized in a disk with external magnetic field of $160\text{mT}$, the frequencies of the circular modes are obtained as $\omega_+/2\pi = 1.12 \text{GHz}$ and $\omega_-/2\pi = -3.40\text{GHz}$, see Supplementary material for details. We then arrive at $M = 1.60 \times 10^{-22}$ $\text{Kg}$ and $\mathcal{K} = 0.024$ $\text{J}/\text{m}^2$. Assuming that the bubble mass mostly depends on its radius, we  use the same mass for the  bubble of the same radius in other geometries.\cite{Buettner.Moutafis.ea:NP2015}

We now consider the effect of strain induced by SAW in a wire geometry. The strain shape factors in Eq.~(\ref{shapefactors}) are estimated using the bubble ansatz, Eq.(\ref{ansatz}).
We consider a relatively long symmetric nanowire of width $200\text{nm}$ and thickness $t = 32\text{nm}$ with perpendicular anisotropy. Throughout the paper we use parameters corresponding to FePt: the saturation magnetization $M_s = 10^6 \text{A}/\text{m}$, the exchange constant $A = 10^{-11}\text{J}/\text{m}$, the perpendicular anisotropy $K_u = 1.3 \times 10^6 \text{J}/\text{m}^3$, and the magnetoelastic constant $B_1 = 6.6 \times 10^6 \text{J}/\text{m}^3$.\cite{Moutafis.Komineas.ea:PRB2009,Spada.Parker.ea:JAP2003}  Gilbert damping $\alpha = 0.1$ is used throughout the paper unless explicitly mentioned. We consider counter-propagating SAWs of frequency $\omega/2\pi = 4.23\text{GHz}$ that travel along the nanowire. For the piezoelectric substrate we use the speed of sound $2114\text{m/s}$ which corresponds to  PZT.\cite{Dean.Bryan.ea:APL2015} In a long nanowire, only the edge repulsion along the transverse direction should be considered which leads to the force corresponding to parabolic potential in the $y$-direction. The edge repulsion along the $y$-direction will suppress the side motion due to the Magnus force. At the same time the external force due to strain gradient along the $x$-direction pushes the bubble horizontally along the nanowire. Since the strain gradient induced by SAWs is periodic with corresponding force vanishing at the anti-nodes of standing wave, the induced force dynamically pushes the bubble towards the anti-nodes. We show this behavior in Fig.~\ref{dynamics} (left). This shows that counter-propagating SAWs form pinning points at the anti-nodes of the standing wave in a nanowire. 

The pinning of a bubble at anti-nodes of standing wave can be utilized to drive the bubble by giving a small velocity to the standing wave through detuning between the counter-propagating waves. When the wave moves with a velocity $v$, for an observer in the moving frame of reference, the change in frequency i.e. detuning $\Delta \omega$, can be obtained using Doppler effect as $\Delta \omega = (v/v_g) \omega$, where $v_g = \omega/k$ is the group velocity of the SAWs in the nanowire. Therefore, the bubble velocity, which is coupled with the standing wave, is directly proportional to the detuning between the counter-propagating waves. However, this relation breaks beyond some critical value of the detuning at which the bubble starts to skip anti-nodes. Then the velocity of the bubble is not proportional to the detuning as shown in Fig.~\ref{dynamics} (middle). For strain of magnitude $\varepsilon_{xx}^0 = 5 \times 10^{-4}$, the bubble velocity increases with detuning and reaches maximum of $1.2\text{cm/s}$ at $\Delta \omega = 1.6\text{MHz}$. On further increasing the detuning,  the bubble decouples from the anti-node of standing wave and no longer moves  along the anti-node.  This results in oscillatory motion of the bubble that interacts with the different regions of the wave with reduced net velocity.

In order to drive a magnetic bubble at a high velocity, a strong coupling of the bubble with the anti-node of counter-propagating SAWs is necessary. Since the strain shape factors are constants for a fixed shape and size of the bubble, stronger coupling can be achieved for higher strain magnitude or larger magnetoelastic constants. Therefore, the maximum bubble velocity for a suitable detuning is directly proportional to the strain magnitude and/or magnetoeleastic constants. For strain magnitude $\varepsilon_{xx}^0 = 10^{-4}$ with magnetoelastic constants mentioned above, we obtain the bubble velocity of about $0.5\text{mm/s}$ at detuning $\Delta \omega = 100\text{KHz}$. Numerical analysis indicates that by careful device engineering order of magnitude larger strain can be created in piezoelectric materials.\cite{Dean.Bryan.ea:APL2015} We can achieve higher bubble velocity of $4.5\text{cm/s}$ at very high strain magnitude of $10^{-3}$ and detuning $5.8\text{MHz}$. Large strain can be also realized by elastic standing waves injected through laser pulse.\cite{PhysRevLett.112.147403, PhysRevB.84.214432} 

Without confinement by the edge, skyrmions or magnetic bubbles with topological charge exhibit Hall-like motion.\cite{Kovalev:PRB2014,PhysRevLett.107.136804, Jiang2016a} This can be achieved in a large magnetic film where the Magnus force becomes the dominant force component driving the skyrmion  bubble along the transverse direction. In Eq.~(\ref{eoms}) above, the restoring force on the bubble corresponding to demagnetizing field from the edge potential is absent, i.e. $\mathcal{K} = 0$. The bubble moves in the longitudinal direction towards the anti-nodes, and there will also be a motion along the transverse direction. Since transverse velocity is larger than longitudinal by a factor of $1/\alpha$, the transverse displacement is more pronounced and it also saturates once the bubble reaches the anti-node. Such transverse displacement could also serve as a manifestation of coupling between the skyrmion bubble and the standing wave from counter-propagating SAWs.

\begin{figure}[htb!]
\vspace{0.5cm}
\centering
\includegraphics[width=1.05\columnwidth]{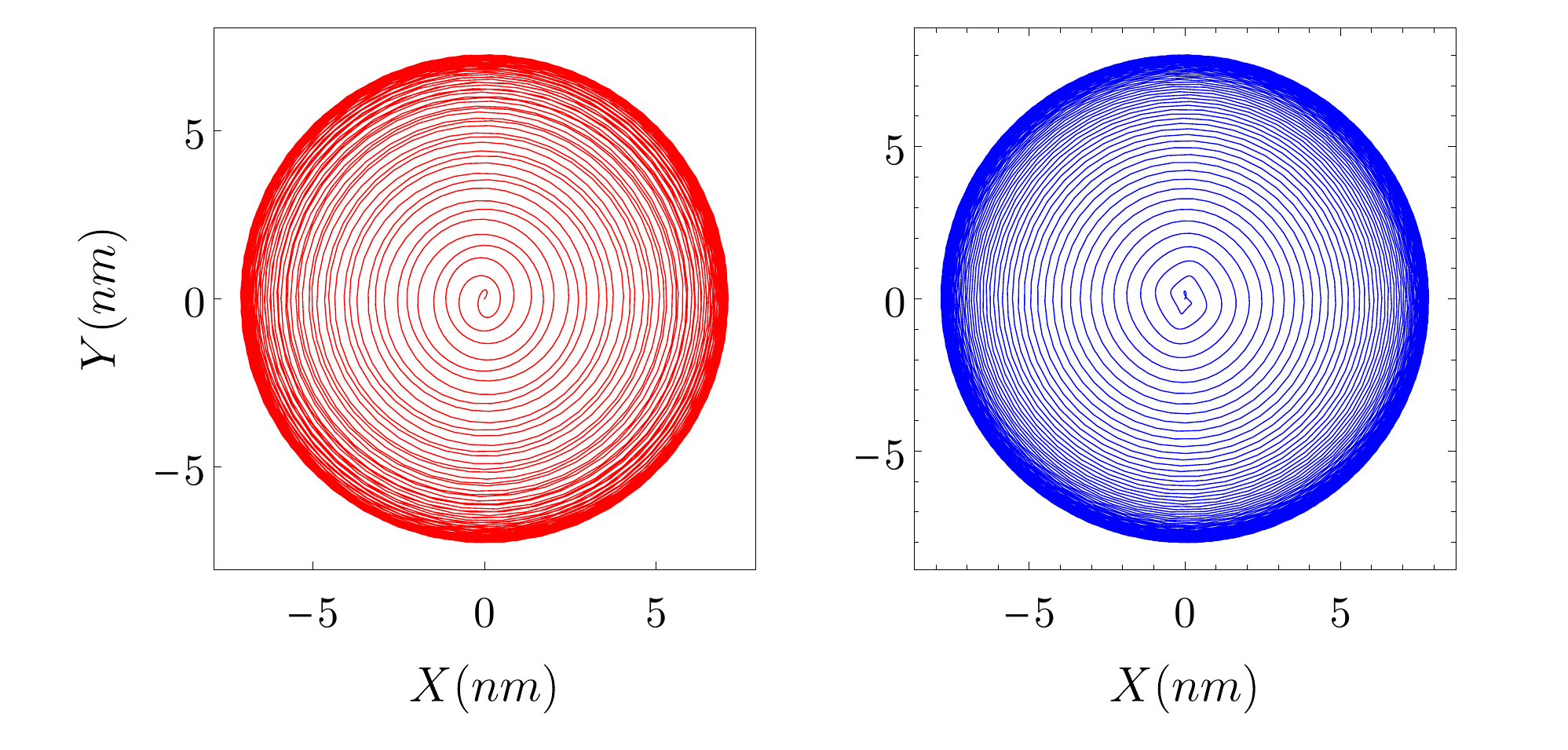}
\caption{(Color online) Resonant rotation of the center of a magnetic bubble in a disk by counter-propagating SAWs of frequencies (left/red) $-3.40\text{GHz}$ and  (right/blue) $1.12\text{GHz}$.}
\label{rotation}
\end{figure}
Finally, we consider SAW induced skyrmion motion in a disk. With parabolic edge potential, the equation of motion in Eq.~(\ref{eoms}) has two circular modes with frequencies $\omega_\pm$, see Eq.(\ref{frequencies}) , in the absence of external strain. If $\omega_-/\omega_+ = -4$, the bubble follows the path of a hypocycloid with five cusps. In a disk of radius $100\text{nm}$, a bubble of radius $52\text{nm}$ follows a not-trivial path with four cusps and can be represented as the superposition of two spirals with eigen-frequencies $\omega_+ = 1.12\text{GHz}$ and $\omega_- = -3.40\text{GHz}$  that travel in opposite directions, see Supplementary material. Here, by exciting each of the eigen-frequencies of free bubble oscillation in the disk, we can realize a resonant bubble oscillator. Similar current induced skyrmion oscillator has been realized by the balance of tangential and radial component of current induced torque in a disk. \cite{1367-2630-17-2-023061} In our study, we apply  counter-propagating SAWs of either of the two resonant frequencies across the disk. The bubble can be resonantly excited, spirally increasing the radius of its path, see Fig.~\ref{rotation}. Once the Gilbert damping losses balance the SAW driving, the bubble reaches the limiting cycle at a maximum radius. We can notice from Fig.~\ref{rotation} that the bubble circulates in the disk in opposite directions depending upon the resonant frequency. For a skyrmion bubble of finite mass moving in a disk of sufficiently large radius, we can estimate the maximum radius of rotation, 
$R_{max} \propto F_{max} /(4\pi s \alpha \eta\omega )$
where $F_{max}$ is the maximum force due to SAWs in Eq.~(\ref{eoms}). The ratio $F_{max}/\omega$ increases as we lower the frequency saturating at some specific value of $\omega$ for a fixed value of strain amplitude.   
In a disk of radius $100\text{nm}$, a bubble reaches the limiting cycle of radius roughly equal to $8\text{nm}$ with counter-propagating SAWs for both resonant frequencies $1.12\text{GHz}$ and $3.40\text{GHz}$, for which we take strain amplitude $\varepsilon^0_{xx} = 5 \times 10^{-4}$ and Gilbert damping $\alpha = 0.01$. As this is a resonant effect, using the material combinations with lower Gilbert damping will reveal stronger effect.\cite{PhysRevB.95.134440} 

To summarize, we have studied dynamical control of a magnetic skyrmion bubble in a ferromagnet using counter-propagating SAWs. We propose a mechanism of SAWs controlled dynamical pinning of magnetic skyrmion bubbles at anti-nodes. We also demonstrate longitudinal driving of a bubble by SAWs in a narrow nanowire where the longitudinal motion is maintained due to pining at SAW anti-nodes and the transverse motion is suppressed by edge repulsion. In a film geometry, a larger transverse velocity can be achieved due to the Magnus force. We have also proposed a resonant bubble oscillator which can be utilized as an efficient magnetic bubble based microwave generator. In this study, a finite mass plays an important role thus the proposed method can be used in order to study the skyrmion bubble mass. Our theory and Eq.~(\ref{eoms}) also apply to skyrmions stabilized by Dzyaloshinskii-Moriya interactions. Note that the skyrmion bubble mass becomes smaller in the presence of Dzyaloshinskii-Moriya interactions which will introduce some changes in the dynamics.\cite{Schuette.Iwasaki.ea:PRB2014,Woo.Song.ea:NC2017}

We thank Shireen Adenwalla for useful discussions. This work was supported by the DOE Early CareerAward DE-SC0014189. The computations were performed utilizing the Holland Computing Center of the University of Nebraska.

\section*{Appendix}
The inertial mass of a magnetic bubble has been calculated numerically and also roughly confirmed by experiment.\cite{Makhfudz.Krueger.ea:PRL2012, Moutafis.Komineas.ea:PRB2009, Buettner.Moutafis.ea:NP2015} Here, we utilize the numerical method similar to the one described in Ref.~\onlinecite{Makhfudz.Krueger.ea:PRL2012} to estimate the mass of a skyrmion magnetic bubble. We micromagnetically simulate the path of a magnetic bubble in a disk and fit the bubble path with that of two super-imposing spirals to find the frequencies of two circular modes corresponding to the bubble motion. Which are related to the mass of the skyrmion bubble and  spring constant of the parabolic edge potential in the disk according  with Eq.(10) in the main text. 

\renewcommand{\thefigure}{A\arabic{figure}}

\setcounter{figure}{0}

\begin{figure}[!htb]
	\vspace{0.5cm}
	\centering
	\includegraphics[width=0.95\columnwidth]{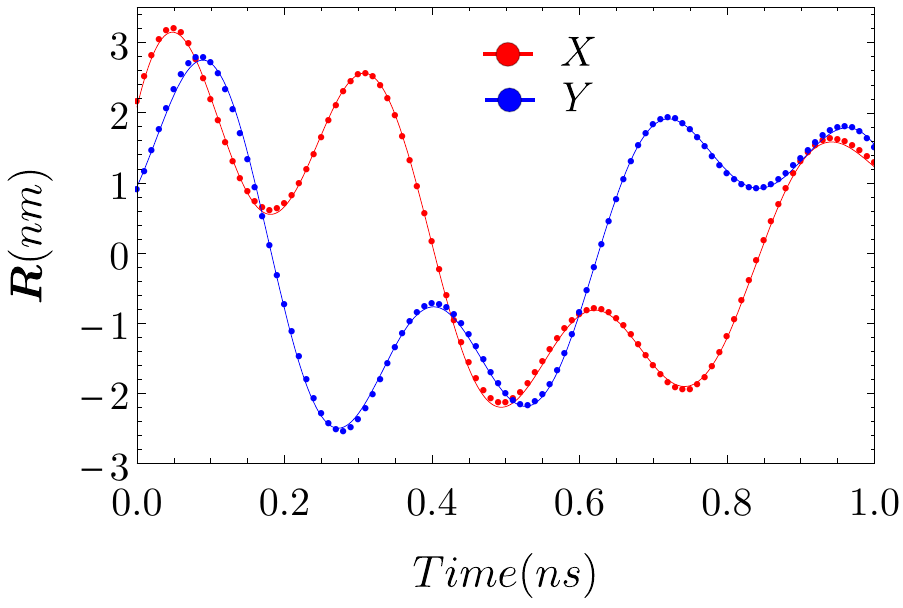}
	\caption{(Color online) The displacement of center of mass of the bubble as a function of time from micromagnetic simulations (points) and best fit to the superposition of two spiral modes (lines).}
	\label{fitting}
\end{figure}

We micromagnetically simulate FePt nanodisk of radius $100 \text{nm}$ and thickness $32 \text{nm}$ using open source micromagnetic simulator mumax3.\cite{Vansteenkiste2014} The magnetic parameters taken are mentioned in the main text. A magnetic bubble of radius $52\text{nm}$ and thickness of circular domain wall $18\text{nm}$ is stabilized. Note that the size of a bubble can be controlled by applying a magnetic field along the direction opposite to the core of the  bubble. 
The bubble at rest is kicked off with a small magnetic field gradient pulse and its trajectroy of the center of mass is recorded. The bubble follows a hypocycloid of four cusps which can be fitted with the superposition of two spirals with eigen-frequencies $\omega_{\pm}$. The fitting of the gyrotropic path of bubble, as shown in Fig.(\ref{fitting}), gives the spiral eigen-frequencies $\omega_+/2\pi = 1.12 \text{GHz}$ and $\omega_-/2\pi = -3.40\text{GHz}$. With these frequencies, we estimate the inertial mass of the magnetic bubble and the stiffness of magnetostatic potential using Eq.(10) as $M = 1.60 \times 10^{-22}$ $\text{Kg}$ and $\mathcal{K} = 0.024$ $\text{J}/\text{m}^2$ respectively. These estimated parameters are used in the calculations done in the main text above.


\begin{thebibliography}{45}
\expandafter\ifx\csname natexlab\endcsname\relax\def\natexlab#1{#1}\fi
\expandafter\ifx\csname bibnamefont\endcsname\relax
  \def\bibnamefont#1{#1}\fi
\expandafter\ifx\csname bibfnamefont\endcsname\relax
  \def\bibfnamefont#1{#1}\fi
\expandafter\ifx\csname citenamefont\endcsname\relax
  \def\citenamefont#1{#1}\fi
\expandafter\ifx\csname url\endcsname\relax
  \def\url#1{\texttt{#1}}\fi
\expandafter\ifx\csname urlprefix\endcsname\relax\def\urlprefix{URL }\fi
\providecommand{\bibinfo}[2]{#2}
\providecommand{\eprint}[2][]{\url{#2}}

\bibitem[{\citenamefont{Boulle et~al.}(2011)\citenamefont{Boulle, Malinowski,
  and Kl{\"a}ui}}]{Boulle.Malinowski.ea:MSaER2011}
\bibinfo{author}{\bibfnamefont{O.}~\bibnamefont{Boulle}},
  \bibinfo{author}{\bibfnamefont{G.}~\bibnamefont{Malinowski}},
  \bibnamefont{and}
  \bibinfo{author}{\bibfnamefont{M.}~\bibnamefont{Kl{\"a}ui}},
  \bibinfo{journal}{Materials Science and Engineering R}
  \textbf{\bibinfo{volume}{72}}, \bibinfo{pages}{159 } (\bibinfo{year}{2011}).

\bibitem[{\citenamefont{Iwasaki et~al.}(2013)\citenamefont{Iwasaki, Mochizuki,
  and Nagaosa}}]{Iwasaki.Mochizuki.ea:NC2013}
\bibinfo{author}{\bibfnamefont{J.}~\bibnamefont{Iwasaki}},
  \bibinfo{author}{\bibfnamefont{M.}~\bibnamefont{Mochizuki}},
  \bibnamefont{and} \bibinfo{author}{\bibfnamefont{N.}~\bibnamefont{Nagaosa}},
  \bibinfo{journal}{Nat. Commun.} \textbf{\bibinfo{volume}{4}},
  \bibinfo{pages}{1463} (\bibinfo{year}{2013}).

\bibitem[{\citenamefont{{Parkin} et~al.}(2008)\citenamefont{{Parkin},
  {Hayashi}, and {Thomas}}}]{Parkin.Hayashi.ea:S2008}
\bibinfo{author}{\bibfnamefont{S.~S.~P.} \bibnamefont{{Parkin}}},
  \bibinfo{author}{\bibfnamefont{M.}~\bibnamefont{{Hayashi}}},
  \bibnamefont{and} \bibinfo{author}{\bibfnamefont{L.}~\bibnamefont{{Thomas}}},
  \bibinfo{journal}{Science} \textbf{\bibinfo{volume}{320}},
  \bibinfo{pages}{190} (\bibinfo{year}{2008}).

\bibitem[{\citenamefont{{Yang} et~al.}(2015)\citenamefont{{Yang}, {Ryu}, and
  {Parkin}}}]{Yang.Ryu.ea:NN2015}
\bibinfo{author}{\bibfnamefont{S.-H.} \bibnamefont{{Yang}}},
  \bibinfo{author}{\bibfnamefont{K.-S.} \bibnamefont{{Ryu}}}, \bibnamefont{and}
  \bibinfo{author}{\bibfnamefont{S.}~\bibnamefont{{Parkin}}},
  \bibinfo{journal}{Nat. Nanotechnol.} \textbf{\bibinfo{volume}{10}},
  \bibinfo{pages}{221} (\bibinfo{year}{2015}).

\bibitem[{\citenamefont{Fert et~al.}(2013)\citenamefont{Fert, Cros, and
  Sampaio}}]{Fert.Cros.ea:NN2013}
\bibinfo{author}{\bibfnamefont{A.}~\bibnamefont{Fert}},
  \bibinfo{author}{\bibfnamefont{V.}~\bibnamefont{Cros}}, \bibnamefont{and}
  \bibinfo{author}{\bibfnamefont{J.}~\bibnamefont{Sampaio}},
  \bibinfo{journal}{Nat. Nanotechnol.} \textbf{\bibinfo{volume}{8}},
  \bibinfo{pages}{152} (\bibinfo{year}{2013}).

\bibitem[{\citenamefont{{Beach} et~al.}(2008)\citenamefont{{Beach}, {Tsoi}, and
  {Erskine}}}]{Beach.Tsoi.ea:JMMM2008}
\bibinfo{author}{\bibfnamefont{G.~S.~D.} \bibnamefont{{Beach}}},
  \bibinfo{author}{\bibfnamefont{M.}~\bibnamefont{{Tsoi}}}, \bibnamefont{and}
  \bibinfo{author}{\bibfnamefont{J.~L.} \bibnamefont{{Erskine}}},
  \bibinfo{journal}{J. Magn. Magn. Mater.} \textbf{\bibinfo{volume}{320}},
  \bibinfo{pages}{1272} (\bibinfo{year}{2008}).

\bibitem[{\citenamefont{{Bogart} et~al.}(2009)\citenamefont{{Bogart},
  {Atkinson}, {O'Shea}, {McGrouther}, and
  {McVitie}}}]{Bogart.Atkinson.ea:PRB2009}
\bibinfo{author}{\bibfnamefont{L.~K.} \bibnamefont{{Bogart}}},
  \bibinfo{author}{\bibfnamefont{D.}~\bibnamefont{{Atkinson}}},
  \bibinfo{author}{\bibfnamefont{K.}~\bibnamefont{{O'Shea}}},
  \bibinfo{author}{\bibfnamefont{D.}~\bibnamefont{{McGrouther}}},
  \bibnamefont{and}
  \bibinfo{author}{\bibfnamefont{S.}~\bibnamefont{{McVitie}}},
  \bibinfo{journal}{Phys. Rev. B} \textbf{\bibinfo{volume}{79}},
  \bibinfo{eid}{054414} (\bibinfo{year}{2009}).

\bibitem[{\citenamefont{Jiang et~al.}(2011)\citenamefont{Jiang, Thomas, Moriya,
  and Parkin}}]{Jiang.Thomas.ea:NL2011}
\bibinfo{author}{\bibfnamefont{X.}~\bibnamefont{Jiang}},
  \bibinfo{author}{\bibfnamefont{L.}~\bibnamefont{Thomas}},
  \bibinfo{author}{\bibfnamefont{R.}~\bibnamefont{Moriya}}, \bibnamefont{and}
  \bibinfo{author}{\bibfnamefont{S.~S.~P.} \bibnamefont{Parkin}},
  \bibinfo{journal}{Nano Lett.} \textbf{\bibinfo{volume}{11}},
  \bibinfo{pages}{96} (\bibinfo{year}{2011}).

\bibitem[{\citenamefont{Reichhardt et~al.}(2015)\citenamefont{Reichhardt, Ray,
  and Reichhardt}}]{Reichhardt.Ray.ea:PRL2015}
\bibinfo{author}{\bibfnamefont{C.}~\bibnamefont{Reichhardt}},
  \bibinfo{author}{\bibfnamefont{D.}~\bibnamefont{Ray}}, \bibnamefont{and}
  \bibinfo{author}{\bibfnamefont{C.~J.~O.} \bibnamefont{Reichhardt}},
  \bibinfo{journal}{Phys. Rev. Lett.} \textbf{\bibinfo{volume}{114}},
  \bibinfo{pages}{217202} (\bibinfo{year}{2015}).

\bibitem[{\citenamefont{Nagaosa and Tokura}(2013)}]{Nagaosa2013}
\bibinfo{author}{\bibfnamefont{N.}~\bibnamefont{Nagaosa}} \bibnamefont{and}
  \bibinfo{author}{\bibfnamefont{Y.}~\bibnamefont{Tokura}},
  \bibinfo{journal}{Nature Nanotechnology} \textbf{\bibinfo{volume}{8}},
  \bibinfo{pages}{899} (\bibinfo{year}{2013}).

\bibitem[{\citenamefont{Kong and Zang}(2013)}]{PhysRevLett.111.067203}
\bibinfo{author}{\bibfnamefont{L.}~\bibnamefont{Kong}} \bibnamefont{and}
  \bibinfo{author}{\bibfnamefont{J.}~\bibnamefont{Zang}},
  \bibinfo{journal}{Phys. Rev. Lett.} \textbf{\bibinfo{volume}{111}},
  \bibinfo{pages}{067203} (\bibinfo{year}{2013}).

\bibitem[{\citenamefont{Lin et~al.}(2014)\citenamefont{Lin, Batista,
  Reichhardt, and Saxena}}]{PhysRevLett.112.187203}
\bibinfo{author}{\bibfnamefont{S.-Z.} \bibnamefont{Lin}},
  \bibinfo{author}{\bibfnamefont{C.~D.} \bibnamefont{Batista}},
  \bibinfo{author}{\bibfnamefont{C.}~\bibnamefont{Reichhardt}},
  \bibnamefont{and} \bibinfo{author}{\bibfnamefont{A.}~\bibnamefont{Saxena}},
  \bibinfo{journal}{Phys. Rev. Lett.} \textbf{\bibinfo{volume}{112}},
  \bibinfo{pages}{187203} (\bibinfo{year}{2014}).

\bibitem[{\citenamefont{Zang et~al.}(2011)\citenamefont{Zang, Mostovoy, Han,
  and Nagaosa}}]{PhysRevLett.107.136804}
\bibinfo{author}{\bibfnamefont{J.}~\bibnamefont{Zang}},
  \bibinfo{author}{\bibfnamefont{M.}~\bibnamefont{Mostovoy}},
  \bibinfo{author}{\bibfnamefont{J.~H.} \bibnamefont{Han}}, \bibnamefont{and}
  \bibinfo{author}{\bibfnamefont{N.}~\bibnamefont{Nagaosa}},
  \bibinfo{journal}{Phys. Rev. Lett.} \textbf{\bibinfo{volume}{107}},
  \bibinfo{pages}{136804} (\bibinfo{year}{2011}).

\bibitem[{\citenamefont{Kovalev}(2014)}]{Kovalev:PRB2014}
\bibinfo{author}{\bibfnamefont{A.~A.} \bibnamefont{Kovalev}},
  \bibinfo{journal}{Phys. Rev. B} \textbf{\bibinfo{volume}{89}},
  \bibinfo{eid}{241101} (\bibinfo{year}{2014}).

\bibitem[{\citenamefont{Woo et~al.}(2016)\citenamefont{Woo, Litzius,
  Kr{\"{u}}ger, Im, Caretta, Richter, Mann, Krone, Reeve, Weigand
  et~al.}}]{Woo2015}
\bibinfo{author}{\bibfnamefont{S.}~\bibnamefont{Woo}},
  \bibinfo{author}{\bibfnamefont{K.}~\bibnamefont{Litzius}},
  \bibinfo{author}{\bibfnamefont{B.}~\bibnamefont{Kr{\"{u}}ger}},
  \bibinfo{author}{\bibfnamefont{M.-y.} \bibnamefont{Im}},
  \bibinfo{author}{\bibfnamefont{L.}~\bibnamefont{Caretta}},
  \bibinfo{author}{\bibfnamefont{K.}~\bibnamefont{Richter}},
  \bibinfo{author}{\bibfnamefont{M.}~\bibnamefont{Mann}},
  \bibinfo{author}{\bibfnamefont{A.}~\bibnamefont{Krone}},
  \bibinfo{author}{\bibfnamefont{R.~M.} \bibnamefont{Reeve}},
  \bibinfo{author}{\bibfnamefont{M.}~\bibnamefont{Weigand}},
  \bibnamefont{et~al.}, \bibinfo{journal}{Nat. Mater.}
  \textbf{\bibinfo{volume}{15}}, \bibinfo{pages}{501} (\bibinfo{year}{2016}).

\bibitem[{\citenamefont{Jiang et~al.}(2017)\citenamefont{Jiang, Zhang, Yu,
  Zhang, Wang, {Benjamin Jungfleisch}, Pearson, Cheng, Heinonen, Wang
  et~al.}}]{Jiang2016a}
\bibinfo{author}{\bibfnamefont{W.}~\bibnamefont{Jiang}},
  \bibinfo{author}{\bibfnamefont{X.}~\bibnamefont{Zhang}},
  \bibinfo{author}{\bibfnamefont{G.}~\bibnamefont{Yu}},
  \bibinfo{author}{\bibfnamefont{W.}~\bibnamefont{Zhang}},
  \bibinfo{author}{\bibfnamefont{X.}~\bibnamefont{Wang}},
  \bibinfo{author}{\bibfnamefont{M.}~\bibnamefont{{Benjamin Jungfleisch}}},
  \bibinfo{author}{\bibfnamefont{J.~E.} \bibnamefont{Pearson}},
  \bibinfo{author}{\bibfnamefont{X.}~\bibnamefont{Cheng}},
  \bibinfo{author}{\bibfnamefont{O.}~\bibnamefont{Heinonen}},
  \bibinfo{author}{\bibfnamefont{K.~L.} \bibnamefont{Wang}},
  \bibnamefont{et~al.}, \bibinfo{journal}{Nat. Phys.}
  \textbf{\bibinfo{volume}{13}}, \bibinfo{pages}{162} (\bibinfo{year}{2017}).

\bibitem[{\citenamefont{Litzius et~al.}(2017)\citenamefont{Litzius, Lemesh,
  Kr{\"u}ger, Bassirian, Caretta, Richter, B{\"u}ttner, Sato, Tretiakov,
  F{\"o}rster et~al.}}]{Litzius.Lemesh.ea:NP2017}
\bibinfo{author}{\bibfnamefont{K.}~\bibnamefont{Litzius}},
  \bibinfo{author}{\bibfnamefont{I.}~\bibnamefont{Lemesh}},
  \bibinfo{author}{\bibfnamefont{B.}~\bibnamefont{Kr{\"u}ger}},
  \bibinfo{author}{\bibfnamefont{P.}~\bibnamefont{Bassirian}},
  \bibinfo{author}{\bibfnamefont{L.}~\bibnamefont{Caretta}},
  \bibinfo{author}{\bibfnamefont{K.}~\bibnamefont{Richter}},
  \bibinfo{author}{\bibfnamefont{F.}~\bibnamefont{B{\"u}ttner}},
  \bibinfo{author}{\bibfnamefont{K.}~\bibnamefont{Sato}},
  \bibinfo{author}{\bibfnamefont{O.~A.} \bibnamefont{Tretiakov}},
  \bibinfo{author}{\bibfnamefont{J.}~\bibnamefont{F{\"o}rster}},
  \bibnamefont{et~al.}, \bibinfo{journal}{Nat. Phys.}
  \textbf{\bibinfo{volume}{13}}, \bibinfo{pages}{170} (\bibinfo{year}{2017}),
  \eprint{1608.07216}.

\bibitem[{\citenamefont{Liang et~al.}(2015)\citenamefont{Liang, Degrave, Stolt,
  Tokura, and Jin}}]{Liang.Degrave.ea:NC2015}
\bibinfo{author}{\bibfnamefont{D.}~\bibnamefont{Liang}},
  \bibinfo{author}{\bibfnamefont{J.~P.} \bibnamefont{Degrave}},
  \bibinfo{author}{\bibfnamefont{M.~J.} \bibnamefont{Stolt}},
  \bibinfo{author}{\bibfnamefont{Y.}~\bibnamefont{Tokura}}, \bibnamefont{and}
  \bibinfo{author}{\bibfnamefont{S.}~\bibnamefont{Jin}}, \bibinfo{journal}{Nat.
  Commun.} \textbf{\bibinfo{volume}{6}}, \bibinfo{eid}{8217}
  (\bibinfo{year}{2015}), \eprint{1503.03523}.

\bibitem[{\citenamefont{Chudnovsky and Jaafar}(2016)}]{PhysRevApplied.5.031002}
\bibinfo{author}{\bibfnamefont{E.~M.} \bibnamefont{Chudnovsky}}
  \bibnamefont{and} \bibinfo{author}{\bibfnamefont{R.}~\bibnamefont{Jaafar}},
  \bibinfo{journal}{Phys. Rev. Applied} \textbf{\bibinfo{volume}{5}},
  \bibinfo{pages}{031002} (\bibinfo{year}{2016}).

\bibitem[{\citenamefont{Kim et~al.}(2016)\citenamefont{Kim, Hill, and
  Tserkovnyak}}]{PhysRevLett.117.237201}
\bibinfo{author}{\bibfnamefont{S.~K.} \bibnamefont{Kim}},
  \bibinfo{author}{\bibfnamefont{D.}~\bibnamefont{Hill}}, \bibnamefont{and}
  \bibinfo{author}{\bibfnamefont{Y.}~\bibnamefont{Tserkovnyak}},
  \bibinfo{journal}{Phys. Rev. Lett.} \textbf{\bibinfo{volume}{117}},
  \bibinfo{pages}{237201} (\bibinfo{year}{2016}).
 

\bibitem[{\citenamefont{Tong et~al.}(2016)\citenamefont{Tong, Fang, Cai, Gong,
  and Duan}}]{Tong.Fang.ea:CMS2016}
\bibinfo{author}{\bibfnamefont{W.-Y.} \bibnamefont{Tong}},
  \bibinfo{author}{\bibfnamefont{Y.-W.} \bibnamefont{Fang}},
  \bibinfo{author}{\bibfnamefont{J.}~\bibnamefont{Cai}},
  \bibinfo{author}{\bibfnamefont{S.-J.} \bibnamefont{Gong}}, \bibnamefont{and}
  \bibinfo{author}{\bibfnamefont{C.-G.} \bibnamefont{Duan}},
  \bibinfo{journal}{Comput. Mater. Sci.} \textbf{\bibinfo{volume}{112}},
  \bibinfo{pages}{467 } (\bibinfo{year}{2016}).

\bibitem[{\citenamefont{Hsu et~al.}(2013)\citenamefont{Hsu, Kubetzka, Finco, Romming, Bergmann  and
		Wiesendanger}}]{Hsu2016}
\bibinfo{author}{\bibfnamefont{Pin-Jui}~\bibnamefont{Hsu}},
\bibinfo{author}{\bibfnamefont{André}~\bibnamefont{Kubetzka}},
\bibinfo{author}{\bibfnamefont{Aurore}~\bibnamefont{Finco}},
\bibinfo{author}{\bibfnamefont{Niklas}~\bibnamefont{Romming}},
\bibinfo{author}{\bibfnamefont{Kirsten~von}~\bibnamefont{Bergmann}},
\bibnamefont{and}
\bibinfo{author}{\bibfnamefont{Roland}~\bibnamefont{Wiesendanger}},
\bibinfo{journal}{Nat. Nanotechnol.} \textbf{\bibinfo{volume}{12}},
\bibinfo{pages}{123} (\bibinfo{year}{2016}).

\bibitem[{\citenamefont{{Upadhyaya} et~al.}(2013)\citenamefont{{Upadhyaya},
  {Dusad}, {Hoffman}, {Tserkovnyak}, {Alzate}, {Amiri}, and
  {Wang}}}]{Upadhyaya.Dusad.ea:PRB2013}
\bibinfo{author}{\bibfnamefont{P.}~\bibnamefont{{Upadhyaya}}},
  \bibinfo{author}{\bibfnamefont{R.}~\bibnamefont{{Dusad}}},
  \bibinfo{author}{\bibfnamefont{S.}~\bibnamefont{{Hoffman}}},
  \bibinfo{author}{\bibfnamefont{Y.}~\bibnamefont{{Tserkovnyak}}},
  \bibinfo{author}{\bibfnamefont{J.~G.} \bibnamefont{{Alzate}}},
  \bibinfo{author}{\bibfnamefont{P.~K.} \bibnamefont{{Amiri}}},
  \bibnamefont{and} \bibinfo{author}{\bibfnamefont{K.~L.}
  \bibnamefont{{Wang}}}, \bibinfo{journal}{Phys. Rev. B}
  \textbf{\bibinfo{volume}{88}}, \bibinfo{eid}{224422} (\bibinfo{year}{2013}).

\bibitem[{\citenamefont{{Belashchenko}
  et~al.}(2016)\citenamefont{{Belashchenko}, {Tchernyshyov}, {Kovalev}, and
  {Tretiakov}}}]{Belashchenko.Tchernyshyov.ea:APL2016}
\bibinfo{author}{\bibfnamefont{K.~D.} \bibnamefont{{Belashchenko}}},
  \bibinfo{author}{\bibfnamefont{O.}~\bibnamefont{{Tchernyshyov}}},
  \bibinfo{author}{\bibfnamefont{A.~A.} \bibnamefont{{Kovalev}}},
  \bibnamefont{and} \bibinfo{author}{\bibfnamefont{O.~A.}
  \bibnamefont{{Tretiakov}}}, \bibinfo{journal}{Appl. Phys. Lett.}
  \textbf{\bibinfo{volume}{108}}, \bibinfo{eid}{132403} (\bibinfo{year}{2016}),
  \eprint{1601.02471}.

\bibitem[{\citenamefont{Davis et~al.}(2010)\citenamefont{Davis, Baruth, and
  Adenwalla}}]{Davis2010}
\bibinfo{author}{\bibfnamefont{S.}~\bibnamefont{Davis}},
  \bibinfo{author}{\bibfnamefont{A.}~\bibnamefont{Baruth}}, \bibnamefont{and}
  \bibinfo{author}{\bibfnamefont{S.}~\bibnamefont{Adenwalla}},
  \bibinfo{journal}{Appl. Phys. Lett.} \textbf{\bibinfo{volume}{97}},
  \bibinfo{pages}{91} (\bibinfo{year}{2010}).

\bibitem[{\citenamefont{{Lei} et~al.}(2013)\citenamefont{{Lei}, {Devolder},
  {Agnus}, {Aubert}, {Daniel}, {Kim}, {Zhao}, {Trypiniotis}, {Cowburn},
  {Chappert} et~al.}}]{Lei.Devolder.ea:NC2013}
\bibinfo{author}{\bibfnamefont{N.}~\bibnamefont{{Lei}}},
  \bibinfo{author}{\bibfnamefont{T.}~\bibnamefont{{Devolder}}},
  \bibinfo{author}{\bibfnamefont{G.}~\bibnamefont{{Agnus}}},
  \bibinfo{author}{\bibfnamefont{P.}~\bibnamefont{{Aubert}}},
  \bibinfo{author}{\bibfnamefont{L.}~\bibnamefont{{Daniel}}},
  \bibinfo{author}{\bibfnamefont{J.-V.} \bibnamefont{{Kim}}},
  \bibinfo{author}{\bibfnamefont{W.}~\bibnamefont{{Zhao}}},
  \bibinfo{author}{\bibfnamefont{T.}~\bibnamefont{{Trypiniotis}}},
  \bibinfo{author}{\bibfnamefont{R.~P.} \bibnamefont{{Cowburn}}},
  \bibinfo{author}{\bibfnamefont{C.}~\bibnamefont{{Chappert}}},
  \bibnamefont{et~al.}, \bibinfo{journal}{Nat. Commun.}
  \textbf{\bibinfo{volume}{4}}, \bibinfo{eid}{1378} (\bibinfo{year}{2013}).

\bibitem[{\citenamefont{{Dean} et~al.}(2015)\citenamefont{{Dean}, {Bryan},
  {Cooper}, {Virbule}, {Cunningham}, and {Hayward}}}]{Dean.Bryan.ea:APL2015}
\bibinfo{author}{\bibfnamefont{J.}~\bibnamefont{{Dean}}},
  \bibinfo{author}{\bibfnamefont{M.~T.} \bibnamefont{{Bryan}}},
  \bibinfo{author}{\bibfnamefont{J.~D.} \bibnamefont{{Cooper}}},
  \bibinfo{author}{\bibfnamefont{A.}~\bibnamefont{{Virbule}}},
  \bibinfo{author}{\bibfnamefont{J.~E.} \bibnamefont{{Cunningham}}},
  \bibnamefont{and} \bibinfo{author}{\bibfnamefont{T.~J.}
  \bibnamefont{{Hayward}}}, \bibinfo{journal}{Appl. Phys. Lett.}
  \textbf{\bibinfo{volume}{107}}, \bibinfo{eid}{142405} (\bibinfo{year}{2015}).


\bibitem[{\citenamefont{{Edrington} et~al.}(2015)\citenamefont{{Edrington}, {Singh},{Dominguez}, {Alexander}, {Nepal}, and {Adenwalla}}}]{Edrington2018}
\bibinfo{author}{\bibfnamefont{Westin}~\bibnamefont{{Edrington}}},
\bibinfo{author}{\bibfnamefont{Uday} \bibnamefont{{Singh}}},
\bibinfo{author}{\bibfnamefont{Maya~Abo} \bibnamefont{{Dominguez}}},
\bibinfo{author}{\bibfnamefont{James~Rehwaldt}~\bibnamefont{{Alexander}}},
\bibinfo{author}{\bibfnamefont{Rabindra} \bibnamefont{{Nepal}}},
\bibnamefont{and} \bibinfo{author}{\bibfnamefont{S.}
	\bibnamefont{{Adenwalla}}}, \bibinfo{journal}{Appl. Phys. Lett.}
\textbf{\bibinfo{volume}{112}}, \bibinfo{eid}{052402} (\bibinfo{year}{2018}).
  
\bibitem[{\citenamefont{{Li} et~al.}(2017)\citenamefont{{Li},
		{Zhang}, {Huang}, {Wang}, {Zhang}, {Liu}, {Zhou}, {Kang}, {Koli} and {Lei}}}]{Li2017}
\bibinfo{author}{\bibfnamefont{Zhi}~\bibnamefont{{Li}}},
\bibinfo{author}{\bibfnamefont{Youguang}~\bibnamefont{{Zhang}}},
\bibinfo{author}{\bibfnamefont{Yanggi}~\bibnamefont{{Huang}}},
\bibinfo{author}{\bibfnamefont{Chengxiang}~\bibnamefont{{Wang}}}, 
\bibinfo{author}{\bibfnamefont{Xichao}~\bibnamefont{{Zhang}}},  
\bibinfo{author}{\bibfnamefont{Yan}~\bibnamefont{{Liu}}},
\bibinfo{author}{\bibfnamefont{Yan}~\bibnamefont{{Zhou}}},
\bibinfo{author}{\bibfnamefont{Wang}~\bibnamefont{{Kang}}},
\bibinfo{author}{\bibfnamefont{Shradha ~Chandrashekhar }~\bibnamefont{{Koli}}},
\bibnamefont{and}
\bibinfo{author}{\bibfnamefont{Na}~\bibnamefont{{Lei}}},
\bibinfo{journal}{J. Magn. Magn. Mater.},
\textbf{\bibinfo{volume}{455}}, \bibinfo{eid}{19-24} (\bibinfo{year}{2017}).

\bibitem[{\citenamefont{{Liang} et~al.}(2015)\citenamefont{{Liang},
  {Sepulveda}, {Hoff}, {Keller}, and {Carman}}}]{Liang.Sepulveda.ea:JAP2015}
\bibinfo{author}{\bibfnamefont{C.-Y.} \bibnamefont{{Liang}}},
  \bibinfo{author}{\bibfnamefont{A.~E.} \bibnamefont{{Sepulveda}}},
  \bibinfo{author}{\bibfnamefont{D.}~\bibnamefont{{Hoff}}},
  \bibinfo{author}{\bibfnamefont{S.~M.} \bibnamefont{{Keller}}},
  \bibnamefont{and} \bibinfo{author}{\bibfnamefont{G.~P.}
  \bibnamefont{{Carman}}}, \bibinfo{journal}{J. Appl. Phys.}
  \textbf{\bibinfo{volume}{118}}, \bibinfo{eid}{174101} (\bibinfo{year}{2015}).

\bibitem[{\citenamefont{{Rousseau} et~al.}(2016)\citenamefont{{Rousseau},
  {Weil}, {Rohart}, and {Mougin}}}]{Rousseau.Weil.ea:SR2016}
\bibinfo{author}{\bibfnamefont{O.}~\bibnamefont{{Rousseau}}},
  \bibinfo{author}{\bibfnamefont{R.}~\bibnamefont{{Weil}}},
  \bibinfo{author}{\bibfnamefont{S.}~\bibnamefont{{Rohart}}}, \bibnamefont{and}
  \bibinfo{author}{\bibfnamefont{A.}~\bibnamefont{{Mougin}}},
  \bibinfo{journal}{Sci. Rep.} \textbf{\bibinfo{volume}{6}},
  \bibinfo{eid}{23038} (\bibinfo{year}{2016}).

\bibitem[{\citenamefont{{D'Souza} et~al.}(2016)\citenamefont{{D'Souza}, {Salehi
  Fashami}, {Bandyopadhyay}, and
  {Atulasimha}}}]{DSouza.SalehiFashami.ea:NL2016}
\bibinfo{author}{\bibfnamefont{N.}~\bibnamefont{{D'Souza}}},
  \bibinfo{author}{\bibfnamefont{M.}~\bibnamefont{{Salehi Fashami}}},
  \bibinfo{author}{\bibfnamefont{S.}~\bibnamefont{{Bandyopadhyay}}},
  \bibnamefont{and}
  \bibinfo{author}{\bibfnamefont{J.}~\bibnamefont{{Atulasimha}}},
  \bibinfo{journal}{Nano Lett.} \textbf{\bibinfo{volume}{16}},
  \bibinfo{pages}{1069} (\bibinfo{year}{2016}).
  
\bibitem[{\citenamefont{{Zhang} et~al.}(2015)\citenamefont{{Rousseau},
		{Zhou}, {Ezawa}, {Zhao} and {Zhao}}}]{Zhang2015}
\bibinfo{author}{\bibfnamefont{Xichao}~\bibnamefont{{Zhang}}},
\bibinfo{author}{\bibfnamefont{Yan}~\bibnamefont{{Zhou}}},
\bibinfo{author}{\bibfnamefont{Motohiko}~\bibnamefont{{Ezawa}}},
\bibinfo{author}{\bibfnamefont{G.~P.}~\bibnamefont{{Zhao}}},   
\bibnamefont{and}
\bibinfo{author}{\bibfnamefont{Weisheng}~\bibnamefont{{Zhao}}},
\bibinfo{journal}{Sci. Rep.} \textbf{\bibinfo{volume}{5}},
\bibinfo{eid}{11369} (\bibinfo{year}{2015}).

\bibitem[{\citenamefont{Makhfudz et~al.}(2012)\citenamefont{Makhfudz,
  Kr{\"{u}}ger, and Tchernyshyov}}]{Makhfudz.Krueger.ea:PRL2012}
\bibinfo{author}{\bibfnamefont{I.}~\bibnamefont{Makhfudz}},
  \bibinfo{author}{\bibfnamefont{B.}~\bibnamefont{Kr{\"{u}}ger}},
  \bibnamefont{and}
  \bibinfo{author}{\bibfnamefont{O.}~\bibnamefont{Tchernyshyov}},
  \bibinfo{journal}{Phys. Rev. Lett.} \textbf{\bibinfo{volume}{109}},
  \bibinfo{pages}{217201} (\bibinfo{year}{2012}).

\bibitem[{\citenamefont{Moon et~al.}(2014)\citenamefont{Moon, Chun, Kim, Qiu,
  and Hwang}}]{Moon.Chun.ea:PRB2014}
\bibinfo{author}{\bibfnamefont{K.-W.} \bibnamefont{Moon}},
  \bibinfo{author}{\bibfnamefont{B.~S.} \bibnamefont{Chun}},
  \bibinfo{author}{\bibfnamefont{W.}~\bibnamefont{Kim}},
  \bibinfo{author}{\bibfnamefont{Z.~Q.} \bibnamefont{Qiu}}, \bibnamefont{and}
  \bibinfo{author}{\bibfnamefont{C.}~\bibnamefont{Hwang}},
  \bibinfo{journal}{Phys. Rev. B} \textbf{\bibinfo{volume}{89}},
  \bibinfo{eid}{064413} (\bibinfo{year}{2014}).

\bibitem[{\citenamefont{B{\"u}ttner et~al.}(2015)\citenamefont{B{\"u}ttner,
  Moutafis, Schneider, Kr{\"u}ger, G{\"u}nther, Geilhufe, Schmising, Mohanty,
  Pfau, Schaffert et~al.}}]{Buettner.Moutafis.ea:NP2015}
\bibinfo{author}{\bibfnamefont{F.}~\bibnamefont{B{\"u}ttner}},
  \bibinfo{author}{\bibfnamefont{C.}~\bibnamefont{Moutafis}},
  \bibinfo{author}{\bibfnamefont{M.}~\bibnamefont{Schneider}},
  \bibinfo{author}{\bibfnamefont{B.}~\bibnamefont{Kr{\"u}ger}},
  \bibinfo{author}{\bibfnamefont{C.~M.} \bibnamefont{G{\"u}nther}},
  \bibinfo{author}{\bibfnamefont{J.}~\bibnamefont{Geilhufe}},
  \bibinfo{author}{\bibfnamefont{C.~V.~K.} \bibnamefont{Schmising}},
  \bibinfo{author}{\bibfnamefont{J.}~\bibnamefont{Mohanty}},
  \bibinfo{author}{\bibfnamefont{B.}~\bibnamefont{Pfau}},
  \bibinfo{author}{\bibfnamefont{S.}~\bibnamefont{Schaffert}},
  \bibnamefont{et~al.}, \bibinfo{journal}{Nat. Phys.}
  \textbf{\bibinfo{volume}{11}}, \bibinfo{pages}{225} (\bibinfo{year}{2015}).

\bibitem[{\citenamefont{Campbell and Jones}(1968)}]{Campbell.Jones:IToSaU1968}
\bibinfo{author}{\bibfnamefont{J.~J.} \bibnamefont{Campbell}} \bibnamefont{and}
  \bibinfo{author}{\bibfnamefont{W.~R.} \bibnamefont{Jones}},
  \bibinfo{journal}{IEEE Transactions on Sonics and Ultrasonics}
  \textbf{\bibinfo{volume}{15}}, \bibinfo{pages}{209} (\bibinfo{year}{1968}).

\bibitem[{\citenamefont{Thiele}(1973)}]{Thiele1973}
\bibinfo{author}{\bibfnamefont{A.~A.} \bibnamefont{Thiele}},
  \bibinfo{journal}{Phys. Rev. Lett.} \textbf{\bibinfo{volume}{30}},
  \bibinfo{pages}{230} (\bibinfo{year}{1973}).

\bibitem[{\citenamefont{Malozemoff and Slonczewski}(1979)}]{Malozemoff}
\bibinfo{author}{\bibfnamefont{A.~P.} \bibnamefont{Malozemoff}}
  \bibnamefont{and} \bibinfo{author}{\bibfnamefont{J.~C.}
  \bibnamefont{Slonczewski}}, \emph{\bibinfo{title}{Magnetic Domain Walls in
  Bubble Materials}} (\bibinfo{publisher}{Academic Press, New York},
  \bibinfo{year}{1979}).

\bibitem[{\citenamefont{Abramowitz}(1974)}]{Abramowitz}
\bibinfo{author}{\bibfnamefont{M.}~\bibnamefont{Abramowitz}},
  \emph{\bibinfo{title}{Handbook of Mathematical Functions, With Formulas,
  Graphs, and Mathematical Tables,}} (\bibinfo{publisher}{Dover Publications,
  Incorporated}, \bibinfo{year}{1974}).

\bibitem[{\citenamefont{Vansteenkiste et~al.}(2014)\citenamefont{Vansteenkiste,
  Leliaert, Dvornik, Helsen, Garcia-Sanchez, and {Van
  Waeyenberge}}}]{Vansteenkiste2014}
\bibinfo{author}{\bibfnamefont{A.}~\bibnamefont{Vansteenkiste}},
  \bibinfo{author}{\bibfnamefont{J.}~\bibnamefont{Leliaert}},
  \bibinfo{author}{\bibfnamefont{M.}~\bibnamefont{Dvornik}},
  \bibinfo{author}{\bibfnamefont{M.}~\bibnamefont{Helsen}},
  \bibinfo{author}{\bibfnamefont{F.}~\bibnamefont{Garcia-Sanchez}},
  \bibnamefont{and} \bibinfo{author}{\bibfnamefont{B.}~\bibnamefont{{Van
  Waeyenberge}}}, \bibinfo{journal}{AIP Adv.} \textbf{\bibinfo{volume}{4}},
  \bibinfo{pages}{107133} (\bibinfo{year}{2014}).

\bibitem[{\citenamefont{Moutafis et~al.}(2009)\citenamefont{Moutafis, Komineas,
  and Bland}}]{Moutafis.Komineas.ea:PRB2009}
\bibinfo{author}{\bibfnamefont{C.}~\bibnamefont{Moutafis}},
  \bibinfo{author}{\bibfnamefont{S.}~\bibnamefont{Komineas}}, \bibnamefont{and}
  \bibinfo{author}{\bibfnamefont{J.~A.~C.} \bibnamefont{Bland}},
  \bibinfo{journal}{Phys. Rev. B} \textbf{\bibinfo{volume}{79}},
  \bibinfo{eid}{224429} (\bibinfo{year}{2009}).

\bibitem[{\citenamefont{Spada et~al.}(2003)\citenamefont{Spada, Parker, Platt,
  and Howard}}]{Spada.Parker.ea:JAP2003}
\bibinfo{author}{\bibfnamefont{F.~E.} \bibnamefont{Spada}},
  \bibinfo{author}{\bibfnamefont{F.~T.} \bibnamefont{Parker}},
  \bibinfo{author}{\bibfnamefont{C.~L.} \bibnamefont{Platt}}, \bibnamefont{and}
  \bibinfo{author}{\bibfnamefont{J.~K.} \bibnamefont{Howard}},
  \bibinfo{journal}{J. Appl. Phys.} \textbf{\bibinfo{volume}{94}},
  \bibinfo{pages}{5123} (\bibinfo{year}{2003}).

\bibitem[{\citenamefont{Afanasiev et~al.}(2014)\citenamefont{Afanasiev,
  Razdolski, Skibinsky, Bolotin, Yagupov, Strugatsky, Kirilyuk, Rasing, and
  Kimel}}]{PhysRevLett.112.147403}
\bibinfo{author}{\bibfnamefont{D.}~\bibnamefont{Afanasiev}},
  \bibinfo{author}{\bibfnamefont{I.}~\bibnamefont{Razdolski}},
  \bibinfo{author}{\bibfnamefont{K.~M.} \bibnamefont{Skibinsky}},
  \bibinfo{author}{\bibfnamefont{D.}~\bibnamefont{Bolotin}},
  \bibinfo{author}{\bibfnamefont{S.~V.} \bibnamefont{Yagupov}},
  \bibinfo{author}{\bibfnamefont{M.~B.} \bibnamefont{Strugatsky}},
  \bibinfo{author}{\bibfnamefont{A.}~\bibnamefont{Kirilyuk}},
  \bibinfo{author}{\bibfnamefont{T.}~\bibnamefont{Rasing}}, \bibnamefont{and}
  \bibinfo{author}{\bibfnamefont{A.~V.} \bibnamefont{Kimel}},
  \bibinfo{journal}{Phys. Rev. Lett.} \textbf{\bibinfo{volume}{112}},
  \bibinfo{pages}{147403} (\bibinfo{year}{2014}).

\bibitem[{\citenamefont{Linnik et~al.}(2011)\citenamefont{Linnik, Scherbakov,
  Yakovlev, Liu, Furdyna, and Bayer}}]{PhysRevB.84.214432}
\bibinfo{author}{\bibfnamefont{T.~L.} \bibnamefont{Linnik}},
  \bibinfo{author}{\bibfnamefont{A.~V.} \bibnamefont{Scherbakov}},
  \bibinfo{author}{\bibfnamefont{D.~R.} \bibnamefont{Yakovlev}},
  \bibinfo{author}{\bibfnamefont{X.}~\bibnamefont{Liu}},
  \bibinfo{author}{\bibfnamefont{J.~K.} \bibnamefont{Furdyna}},
  \bibnamefont{and} \bibinfo{author}{\bibfnamefont{M.}~\bibnamefont{Bayer}},
  \bibinfo{journal}{Phys. Rev. B} \textbf{\bibinfo{volume}{84}},
  \bibinfo{pages}{214432} (\bibinfo{year}{2011}).

\bibitem[{\citenamefont{Zhang et~al.}(2015)\citenamefont{Zhang, Wang, Zheng,
  Zhu, Liu, Chen, Jin, Liu, Jia, and Xue}}]{1367-2630-17-2-023061}
\bibinfo{author}{\bibfnamefont{S.}~\bibnamefont{Zhang}},
  \bibinfo{author}{\bibfnamefont{J.}~\bibnamefont{Wang}},
  \bibinfo{author}{\bibfnamefont{Q.}~\bibnamefont{Zheng}},
  \bibinfo{author}{\bibfnamefont{Q.}~\bibnamefont{Zhu}},
  \bibinfo{author}{\bibfnamefont{X.}~\bibnamefont{Liu}},
  \bibinfo{author}{\bibfnamefont{S.}~\bibnamefont{Chen}},
  \bibinfo{author}{\bibfnamefont{C.}~\bibnamefont{Jin}},
  \bibinfo{author}{\bibfnamefont{Q.}~\bibnamefont{Liu}},
  \bibinfo{author}{\bibfnamefont{C.}~\bibnamefont{Jia}}, \bibnamefont{and}
  \bibinfo{author}{\bibfnamefont{D.}~\bibnamefont{Xue}}, \bibinfo{journal}{New
  Journal of Physics} \textbf{\bibinfo{volume}{17}}, \bibinfo{pages}{023061}
  (\bibinfo{year}{2015}).

\bibitem[{\citenamefont{Barati and Cinal}(2017)}]{PhysRevB.95.134440}
\bibinfo{author}{\bibfnamefont{E.}~\bibnamefont{Barati}} \bibnamefont{and}
  \bibinfo{author}{\bibfnamefont{M.}~\bibnamefont{Cinal}},
  \bibinfo{journal}{Phys. Rev. B} \textbf{\bibinfo{volume}{95}},
  \bibinfo{pages}{134440} (\bibinfo{year}{2017}).

\bibitem[{\citenamefont{Sch{\"u}tte et~al.}(2014)\citenamefont{Sch{\"u}tte,
  Iwasaki, Rosch, and Nagaosa}}]{Schuette.Iwasaki.ea:PRB2014}
\bibinfo{author}{\bibfnamefont{C.}~\bibnamefont{Sch{\"u}tte}},
  \bibinfo{author}{\bibfnamefont{J.}~\bibnamefont{Iwasaki}},
  \bibinfo{author}{\bibfnamefont{A.}~\bibnamefont{Rosch}}, \bibnamefont{and}
  \bibinfo{author}{\bibfnamefont{N.}~\bibnamefont{Nagaosa}},
  \bibinfo{journal}{Phys. Rev. B} \textbf{\bibinfo{volume}{90}},
  \bibinfo{eid}{174434} (\bibinfo{year}{2014}).

\bibitem[{\citenamefont{Woo et~al.}(2017)\citenamefont{Woo, Song, Han, Jung,
  Im, Lee, Song, Fischer, Hong, Choi et~al.}}]{Woo.Song.ea:NC2017}
\bibinfo{author}{\bibfnamefont{S.}~\bibnamefont{Woo}},
  \bibinfo{author}{\bibfnamefont{K.~M.} \bibnamefont{Song}},
  \bibinfo{author}{\bibfnamefont{H.-S.} \bibnamefont{Han}},
  \bibinfo{author}{\bibfnamefont{M.-S.} \bibnamefont{Jung}},
  \bibinfo{author}{\bibfnamefont{M.-Y.} \bibnamefont{Im}},
  \bibinfo{author}{\bibfnamefont{K.-S.} \bibnamefont{Lee}},
  \bibinfo{author}{\bibfnamefont{K.~S.} \bibnamefont{Song}},
  \bibinfo{author}{\bibfnamefont{P.}~\bibnamefont{Fischer}},
  \bibinfo{author}{\bibfnamefont{J.-I.} \bibnamefont{Hong}},
  \bibinfo{author}{\bibfnamefont{J.~W.} \bibnamefont{Choi}},
  \bibnamefont{et~al.}, \bibinfo{journal}{Nat. Commun.}
  \textbf{\bibinfo{volume}{8}}, \bibinfo{eid}{15573} (\bibinfo{year}{2017}).

\end{thebibliography}
\end{document}